\documentstyle[12pt,aasms4]{article}
\begin{document} 

\received{}
\accepted{}
\journalid{}{}
\articleid{}{}

\slugcomment{}
\lefthead{}
\righthead{}

\newcommand{\be}{\begin{equation}}
\newcommand{\en}{\end{equation}}

\def\ltsima{$\; \buildrel < \over \sim \;$}
\def\lsim{\lower.5ex\hbox{\ltsima}}
\def\gtsima{$\; \buildrel > \over \sim \;$}
\def\gsim{\lower.5ex\hbox{\gtsima}}
\def\spose#1{\hbox to 0pt{#1\hss}}
\def\approxlt{\mathrel{\spose{\lower 3pt\hbox{$\sim$}}
        \raise 2.0pt\hbox{$<$}}}
\def\approxgt{\mathrel{\spose{\lower 3pt\hbox{$\sim$}}
        \raise 2.0pt\hbox{$>$}}}
\def\deg {^\circ}
\def\mdot {\dot M}
\def\kms {$\sim$km$\sim$s$^{-1}$}
\def\gs {$\sim$g$\sim$s$^{-1}$}
\def\ergs {$\sim$erg$\sim$s$^{-1}$}
\def\cmtre {$\sim$cm$^{-3}$}\def\nupa{\vfill\eject\noindent}
\def\der#1#2{{d #1 \over d #2}}
\def\l#1{\lambda_{#1}}
\def\grb{$\gamma$-ray burst}
\def\grbs{$\gamma$-ray bursts}
\def\rosat{{\sl ROSAT} }
\def\cmdue {cm$^{-2}$}
\def\gcm {$\sim$g$\sim$cm$^{-3}$}
\def\rsole{$\sim$R_{\odot}}
\def\msole{$\sim$M_{\odot}}
\def\aa #1 #2 {A\&A, {#1}, #2}
\def\mon #1 #2 {MNRAS, {#1}, #2}
\def\apj #1 #2 {ApJ, {#1}, #2}
\def\nat #1 #2 {Nature, {#1}, #2}
\def\pasj #1 #2 {PASJ, {#1}, #2}
\newfont{\mc}{cmcsc10 scaled\magstep2}
\newfont{\cmc}{cmcsc10 scaled\magstep1}
\newcommand{\bc}{\begin{center}}
\newcommand{\ec}{\end{center}}

\title{The Zoo of X--ray sources in the Galactic Center Region:\\
     Observations  with  $BeppoSAX$}

\author{L.Sidoli\altaffilmark{1,2},
		S.Mereghetti\altaffilmark{1}, 
		G.L.Israel\altaffilmark{3,4},
		L.Chiappetti\altaffilmark{1}, 
		A.Treves\altaffilmark{5} and 
		M.Orlandini\altaffilmark{6}}
 
\altaffiltext{1}{Istituto di Fisica Cosmica ``G.Occhialini", C.N.R., via Bassini 15, 20133 Milano, Italy. } 
\altaffiltext{2}{Universit\`a di Milano, Sez. Astrofisica, via Celoria 16, 20133 Milano, Italy. } 
\altaffiltext{3}{Osservatorio Astronomico di Roma, Via Frascati 33, Monteporzio Catone, Roma, Italy.}
\altaffiltext{4}{Affiliated to I.C.R.A.}
\altaffiltext{5}{Universit\`a degli Studi dell'Insubria, Polo di Como, Dipartimento di
Scienze Chimiche, Fisiche e Matematiche, Via Lucini 3, 22100, Como, Italy.}
\altaffiltext{6}{TeSRE, via Gobetti 101, 40129 Bologna, Italy.}
 
\begin{abstract}
  
We report the results of a survey of the  Galactic Center region
($\mid l\mid < 2\deg$, $\mid b\mid < 0.5\deg$) performed
with  the $BeppoSAX$ satellite.
The flux from the center of  our Galaxy 
corresponds to a luminosity of  $\sim$3 10$^{35}$
erg s$^{-1}$ in the 2--10 keV range. 
Due to the limited angular resolution ($\gsim1'$)  only part of it
is supposed to come from  Sagittarius~A*, the non--thermal
radio source which is believed to mark the dynamical center 
of the Galaxy.
In addition to the diffuse emission, several bright (L$_X\gsim10^{36}$~ergs~s$^{-1}$)
point sources have been observed, both persistent   
(A~1742--294, SLX~1744--299, SLX~1744--300, 1E~1743.1--2843, 1E~1740.7--2942)
and transient   (XTE~J1748--288, SAX~J1747.0--2853 and KS~1741--293).

The Low Mass X--ray Binary  AX J1745.6--2901, discovered with ASCA at 
only $1.3'$  from SgrA* was detected in a low  luminosity state in August 1997.

The 1--150  keV spectrum of the hard X--ray source   1E~1740.7--2942 is
well described by  a Comptonization model, typical of black hole candidates
in their low/hard state, with no evidence for strong Fe lines.
  
The detection of a type I burst shows that   the transient source  
SAX~J1747.0--2853 (probably the same as the
1976 transient GX 0.2--0.2) is a LMXRB containing a neutron star.

The transient black hole candidate XTE~J1748--288 was detected
at a luminosity   ($\sim 10^{36}$~ergs~s$^{-1}$) consistent
with the extrapolation of the exponential decay of the outburst
observed with the XTE All Sky Monitor.  
  
Two fainter sources are very likely associated with young neutron 
stars: the (possibly diffuse) X--ray source at the center of the
composite supernova remnant G0.9+0.1, and the ``head" of the axially symmetric
radio source G359.23--0.92. The latter has been detected   
above $\sim$ 6 keV, supporting a non--thermal emission mechanism.

\end{abstract}

\keywords{Galaxy: center --- X-rays: individual (A~1742--294, AX~J1745.6--2901,
SLX~1744--299, SLX~1744--300, 1E~1743.1--2843, 
1E~1740.7--2942, XTE~J1748--288, SAX~J1747.0--2853, KS~1741--293, GX~0.2--0.2, 
G0.9+0.1, G359.23--0.92)}

\section{Introduction}
 
The Galactic Center (hereafter GC) region has 
been observed at different spatial scales and energy bands 
by many X--ray missions.

The main results in the 0.5--10 keV range   have been obtained with
the Einstein Observatory (Watson et al., 1981), ROSAT (Predehl \& Trumper, 1994) 
and, more recently, with ASCA (Maeda et al. 1996, 1998; Koyama et al. 1996).
At higher energies, 
the telescopes ART--P (Pavlinsky, Grebenev \& Sunyaev 1994) 
and SIGMA (Goldwurm et al. 1994) on--board the GRANAT satellite performed a monitoring 
of the GC region leading to the discovery 
of many new   sources.
These observations 
have shown that the GC X--ray emission comes 
both from point sources
and from a diffuse component. Within a few degrees 
from the direction of the GC region
there is a strong concentration of X--ray sources. The zoo 
of these objects 
is very rich and comprises  both transient and 
permanent sources, both ``standard" and very peculiar
objects, like e.g. the ``bursting pulsar'' GRO~J1744--28 
(Lewin et al. 1996), the ``microquasar'' 1E~1740.7--2942 (Mirabel et al. 1992),
and the 2 ms  pulsar SAX J1808.4--3658 (in't Zand et al. 1998a, 
Wijnands \& van der Klis  1998).
 
The region 
($|l|<2\deg$)$\times$($|b|<0.5\deg$) 
around the  GC
was observed with the $BeppoSAX$ satellite   
during 1996--1998 for a total of $\sim$140 hours of
effective  exposure time (Sidoli et al., 1998a, 1999).

In this paper we report on the  observations of
the X--ray point sources in the GC region obtained with the  $BeppoSAX$ Narrow 
Field Instruments
(for the $BeppoSAX$ Wide Field Camera results
see, e.g., Ubertini et al. (1999)).
Although our observations cannot equal the spatial resolution 
obtained with ROSAT at lower energies, they provide good images with 
adequate spectral resolution above a few keV. 
The sensitivity is similar to that obtained with ASCA,  that 
thanks to its solid state detectors has a 
higher  spectral resolution. 
On the other hand, the BeppoSAX mirrors provide a narrower and 
more regular point response
function with respect to that of the ASCA telescopes.  
This allows an accurate analysis of the spatial morphology
of this complex sky region and  
reduces  the problem  of stray light contamination from strong
sources outside the field of view.  
The detailed results on 
the X--ray diffuse emission from SgrA East and West 
and from the   molecular
clouds in this region (SgrB2, SgrC, SgrD) will be
reported in a separate paper.

\section{Observations}
  
The $BeppoSAX$ satellite (Boella et al. 1997a) carries a complement of
several imaging and non-imaging X--ray detectors, covering a broad
energy range from 0.1 keV to 300 keV.
Most of the results presented here have been obtained with the imaging 
instruments placed in the focal planes of four identical X--ray Mirror Units:  
the Medium--Energy Concentrator Spectrometer
(MECS, Boella et al. 1997b) and the 
Low--Energy Concentrator Spectrometer (LECS, Parmar et al. 1997).
Each Mirror Unit consists of 30 nested, confocal mirrors with a double
cone approximation to the Wolter I geometry  and a focal length of 185 cm.

The MECS instrument is based on three position--sensitive 
gas--scintillation proportional counters, providing 1.3--10 keV images 
over a field of view with  $\sim28'$ radius.
The MECS is characterized by a total
(three telescopes) effective area of  $\sim$150~cm$^2$ at 6 keV,
a relatively good angular resolution
(50\% power radius of  $\sim75''$ at 6 keV, on-axis) and 
a moderate energy resolution (FWHM $\sim$8.5$\sqrt{6/{\rm E_{keV}}}$\%).
One (M1) of the three nearly identical units composing the MECS had a failure
in May 1997; all the observations performed after that date were carried out
only with the remaining two units (M2 and M3). 
 
The fourth Mirror Unit is associated with the LECS instrument. 
This consists of a  position--sensitive 
gas--scintillation proportional counter utilizing an  
ultra thin, organic detector entrance window (1.25 $\mu$m) and a driftless configuration
to extend the low energy response down to 0.1 keV.
Its spatial and energy resolutions are similar to those of the MECS,
but the field of view is slightly smaller ($\sim18'$ radius). 
Due to UV contamination leaking through the entrance window during
sunlit periods, 
the LECS is operated only at satellite night time,
resulting in effective exposure times smaller than those of the other
$BeppoSAX$ instruments (typically $\sim$50\% of the MECS values).  

For the analysis of 1E~1740.7--2942, which has a hard spectrum 
extending above the MECS energy range
we have also used the  Phoswich Detector System 
 instrument 
(PDS, Frontera et al. 1997).
This non-imaging instrument   points in the same direction of the 
MECS and LECS telescopes and
covers the  15--300 keV energy range. 
It consists of a square array of
four indipendent NaI(Tl)/CsI(Na) phoswich scintillation
detectors equipped with collimators defining a field of view 
of $\sim$1.3$\deg$ (FWHM).
There are two separate collimators, one for each pair of detectors,
that can be independently rocked off-axis to monitor the background.
During our observations the default rocking mode with offset angles of $\pm210'$
and dwell time 96 s was used.
The PDS provides an energy resolution of  
$\sim$15\% at 60 keV (FWHM).

The log of our $BeppoSAX$ observations of the GC region
performed in 1996--98 is shown in Table~1.
The pointing directions were chosen to cover a $\sim4\deg$   long region 
at  b=0$\deg$,
as well as a few known  sources slightly outside the galactic plane.
We chose to 
partially overlap adjacent pointings, in order to continuously 
cover the galactic plane with the central 
part of the MECS field of view ($\sim8'$ radius),
where the instrument sensitivity and angular resolution are best.
As shown in Table 1, a few pointing directions were observed twice,
at time intervals of 6 or 12 months. This was partly planned, 
to study the long term variability of the sources, and partly
resulted from operational constraints. 
The exposure time achieved in the different regions is not uniform,
ranging from a minimum of 7 ks to $\sim$340 ks at the position of 1E~1743.1--2843.
 
An exposure-corrected   mosaic of the MECS images in the  2--10 keV  
energy range  is shown in Fig.~1. 
 
 
\placetable{tbl-1}

\section{Data Analysis}

Except where noted, the spectral analysis of all the point sources 
reported here has been carried out as follows:
the LECS and MECS counts have been extracted from a circular region
with    $4'$  radius centered at the position of the source  and rebinned 
in order to have at least 20 counts per energy bin. 
The $4'$ radius corresponds to an encircled energy of 90\% .
In several  cases we used a smaller extraction radius 
($2'$, $\sim$60\% enclosed energy)
in order to reduce the problems due  to  source crowding or
unfavourable locations in the MECS field of view.

In any case the  response matrices
appropriate to the adopted extraction radius and source position on the 
detector  have been used.
While these   matrices  include the elements
to properly derive  the correct spectral shape of the sources observed off-axis,
the resulting fluxes can be affected by an uncertainty 
in their absolute normalization that in some cases could reach $\sim$40\%.
 
From each spectrum we have subtracted a local background,
extracted from a source free region of the same observation:
in fact, the standard $BeppoSAX$ background spectra
obtained from blank field observations
underestimate the actual background present in the GC region, which is dominated 
by the galactic diffuse emission.
 
The PDS data  were reduced using the 
SAXDAS software package (Version 1.2.0). The time intervals before and
after Earth occultations were excluded and the spurious spikes (Guainazzi \&
Matteuzzi 1997) filtered out.  The background subtraction was performed
with the standard rocking collimator technique.  
The  background-subtracted spectra were rebinned in order to
have at least 100 counts for each energy channel and then analyzed using the
response matrix released on August 31, 1997.

The resulting LECS/MECS/PDS spectra   have been 
fitted with different models implemented in  XSPEC (Version 10.00).
We took into account the known intercalibration normalizations  
between the
different instruments on--board $BeppoSAX$ (Fiore, Guainazzi \& Grandi 1999), 
using different  relative normalizations 
A$_{LECS}\sim$0.85, A$_{MECS}$=1   and  A$_{PDS}\sim$0.8. 
MECS spectra were extracted in the 2--10 keV energy range, while for the
LECS we used only the data below 2 keV 
(we verified that using the LECS  data above this 
energy  did not significantly
improve the results obtained with the MECS).

\section{Results}

Several previously known sources have been observed in our  
$BeppoSAX$ pointings:
1E~1743.1--2843 (Watson et al. 1981),  
the   persistent black hole candidate  1E~1740.7--2942
(Sunyaev et al. 1991a),
the  X--ray bursters SLX~1744--299, SLX~1744--300, A~1742--294, KS~1741--293 
(Skinner et al. 1990; Pavlinsky, Grebenev \& Sunyaev  1994; Kawai et al. 1988),
and the source(s) at the GC position
(Watson et al. 1981). 
We also detected two X--ray transients discovered very recently: 
SAX~J1747.0--2853 (in't Zand et al. 1998b) and XTE~J1748--288 
(Smith, Levine \& Wood 1998), and  
 discovered a new X--ray source that we identified as the plerionic
emission from the composite supernova remnant G0.9+0.1 (Mereghetti,
Sidoli \& Israel 1998).

All the  sources observed in our survey are listed in Table~2.
Note that in Table~2 we have reported the best source positions 
and corresponding uncertainties
available in the literature.
The positions  derived from our analysis have typical
uncertainties of the order of $\sim1'$ 
(95\% confidence level) and  are in agreement with those
of Table~2. 
We have listed in Table 2 also the positions of a few other
objects which are relevant in the following discussion
of the sources near SgrA*.
 
The results for the individual sources are reported in the  following
sections, while in Table~3 we have summarized 
the parameters of the best fits.  
All the luminosities are  given for an assumed
distance of 8.5 kpc, unless stated differently.
 
\placetable{tbl-2}

\placetable{tbl-3}

\subsection{The source(s) at the Galactic Center position (1E~1742.5--2859)}

The GC field was observed for 100 ks in August 1997 (Obs. n.3).
Strong X--ray emission peaked  at the  GC position 
is clearly detected with the LECS and MECS instruments.
The central part of the MECS image  is shown in Fig.~2, where the positions
of   the X--ray sources previously detected with other satellites have been
indicated.
The background subtracted radial profiles of the X--ray emission in the
soft (2--5 keV) and hard  (5--10 keV) energy ranges
are shown in Fig.~3. 
The data points represent the surface brightness measured in concentric
rings centered at the position of SgrA*, while the solid
lines show for comparison
the profiles expected from a single unresolved point source.
The expected profiles have
been normalized to yield the same number of 
counts within 10$'$ as the measured data.
It is clear that the shape of the observed profiles 
is incompatible with 
a single point--like   source. 
The observed radial profile results from the contribution of
a diffuse component  and one (or possibly more) sources close
to the SgrA* position.

\placefigure{fig2}

\placefigure{fig3}

The X--ray images of this region  with the best angular resolution available so far are those
obtained with the ROSAT PSPC instrument
(Predehl \& Trumper, 1994) which, in addition to the diffuse emission,
showed the presence of three different
sources. Their error regions (20$''$ radius) are indicated in Fig.~2 with the 
small circles labelled PT6, PT7 and PT8. 
One of these sources (PT7)   
is  highly absorbed and located within $10''$ from Sgr A*, while
PT6 has been interpreted as a foreground star, due to its smaller
column density.
It is evident that the emission detected with $BeppoSAX$ is peaked 
at the position of PT7, although we cannot exclude that also the other 
ROSAT sources (and AX~J1745.6--2901, see below) contribute to some
of the detected X--rays. 
 
Though the limited angular resolution hampers a detailed analysis
and introduces unavoidable uncertainties, we proceeded as follows to 
estimate the   spectrum and  flux to be ascribed to
the GC point source(s).

A   spectrum was  extracted from a circular region with $2'$ radius 
centered at the position of the X--ray peak flux (see Fig.~2). 
The background was taken from an external annular region ($6'-8'$), in order
to subtract the contribution from the  diffuse emission.

A  power law 
does not give an appropriate fit to the resulting spectrum, which clearly
shows evidence for emission lines at 6.7 keV and also   at lower energies,
indicative
of a possible thermal origin of the emission. 
The sulfur line at $\sim2.4$ keV is  particularly bright.
A power law with a gaussian line   at 6.7 keV gives a better result, with 
a photon index 
$\Gamma\sim2.6$ and a   column density 
$N_H\sim8\times 10^{22}$\cmdue ($\chi^2=1.1$). 
When the centroid energy and width of the gaussian are let free in the fit,
we obtain the values E=6.7 keV,  $\sigma$=13 eV and an equivalent width 
EW$\sim1.2$ keV.
The same results were found fixing the $\sigma$ of 
the iron line at 0. No evidence for a 6.4 keV line of
fluorescent origin is present, contrary to what is found in the
surrounding diffuse emission (a detailed study of which is deferred to a
separate paper).
The 2--10 keV flux corrected for the absorption is 
$\sim4\times 10^{-11}$ ergs \cmdue s$^{-1}$ and the
luminosity is $\sim3\times 10^{35}$  ergs s$^{-1}$  at a distance of 8.5 kpc.

\placefigure{fig4}

We also fitted the data with a thermal plasma model (MEKAL in XSPEC, Mewe, 
Gronenschild  \& van den Oord 1985). It properly 
accounts for the 6.7 keV iron emission line, but leaves some
residuals around 2.4 keV (see Fig.~4), possibly indicating the presence of 
a multitemperature 
plasma.
In order to properly account for the emission lines at lower
energies, we fitted the spectrum with the sum of two   MEKAL components with
different temperatures (kT$_{M1}\sim0.8$ keV and kT$_{M2}\sim5$ keV),
obtaining a higher N$_{H}$.
Also a single MEKAL model (kT$_{M}\sim1.3$ keV) plus a power law with
a gaussian added at 6.7 keV fit well the data.
All the spectral results are summarized in Table~3.

\subsection{AX J1745.6--2901}

During observations of the GC region performed
with the ASCA satellite in 1993 and 1994 
a possibly new source was discovered $\sim1.3'$ SW 
of SgrA* (Maeda et al. 1996). 
The spectrum of this source, later named   AX~J1745.6--2901 (Maeda et al. 1998),
was harder than that of the X--ray emission
coming from the position of Sgr~A*. 
AX~J1745.6--2901 varied by a factor $\sim$5 in flux between
the two   oservations separated by one year, and it showed an X--ray burst 
and periodic (8.4 hr) eclipses (Maeda et al. 1996).
These authors proposed that AX~J1745.6--2901  could be the
quiescent counterpart of the  bright soft X--ray transient 
A1742-289 observed in outburst in 1975 (Eyles et al. 1975, 
Branduardi et al. 1976).
However, the absence of eclipses at the 8.4 hr period in the 
Ariel V  archival data of the A1742-289 outburst (Kennea \& Skinner 1996),
makes this association uncertain. It is therefore likely that
AX~J1745.6--2901 is a different neutron star LMXRB.
 
In order to search for the possible presence of  AX~J1745.6--2901  
in our observations, we compared the MECS images accumulated in the
soft (2--6 keV) and hard (6--10 keV) energy ranges. 
This was done independently for the two MECS units (M2 and M3)
to obtain the best angular resolution by avoiding   possible small 
misalignement errors. 
Furthermore, we selected only time intervals in which the $BeppoSAX$
attitude was determined with two star trackers 
(indeed small deviations $\lsim2'$ may be present 
when no star tracker is active or when only one is in use).

\placefigure{fig5}
 
The resulting images for the M3 unit are shown in Fig.~5.
In the high energy image, an excess emission in the SW
direction is   clearly resolved from the softer (and partly diffuse)
emission  due to the 1E~1742.5--2859 (GC) source(s).
The  position of this excess is consistent with 
that of   AX~J1745.6--2901.
It is visible only
in the higher  energy channels  of M3: in fact the higher background 
due to misplaced events from the iron calibration sources prevents 
its detection also in M2 (Chiappetti et al. 1998).

The weakness of the source and the presence of the nearby much brighter
X--ray emission hampers
an accurate determimation of its flux and  spectrum,
which, in any case,  is consistent  with the lower value observed
with ASCA in 1993 (Maeda et al. 1996).

\subsection{1E~1740.7--2942}

We obtained two observations (Obs. n. 1 and 6) pointed on the 
``microquasar'' source 1E~1740.7--2942 (Mirabel et al. 1992).
Our spectral analysis  is based on the longer  observation performed
in September 1997. 

As shown by previous observations with coded mask imaging 
instruments in the hard X--ray band (Skinner et al. 1987,
Goldwurm et al. 1994), 1E~1740.7--2942 is the strongest persistent source at
energies greater than $\sim$30 keV within a few degrees from the
GC. We therefore analyzed also the PDS data, in which
the source was clearly detected up to $\sim$150 keV. Due to the crowding of 
this region of the sky, particular care was
devoted to the analysis of this non-imaging detector. 
The PDS field of view includes the positions of the 
persistent LMXRB A~1742--294 and of the transient KS~1741--293. 
As shown by our simultaneous MECS data, the latter source   was not active 
at the time of   observation. 
A~1742--294 lies at $\sim30'$  from 1E~1740.7--2942 and it has
a similar  flux in the 2--10 keV range. However,  
 it has a much softer spectrum (see Table~3).
Using our best fit thermal bremsstrahlung results and taking into account 
the reduced off-axis response of the PDS collimator, we estimate
that A~1742--294   contributes less than 10\% to the observed 
counts above 40 keV. 
A further problem arised by the contamination of one of 
the two offset collimator positions used in the background determination  
by the  bright X--ray source GX~3+1. 
We therefore estimated the PDS background using  only   one of the two 
off-axis collimator positions.

A single power law  provides an adequate fit to the 
LECS and MECS spectrum below 10 keV (photon index
$\Gamma\sim$1.5,   $N_{H}\sim$1.5$\times$10$^{23}$ cm$^{-2}$).
We searched in the MECS data for the possible presence  of iron lines
obtaining a negative result: for lines in the range 6.4--6.7 keV 
we
can give a 90\% upper limit to the equivalent width  of $\sim30$ eV 
(with a line width $\sigma$ fixed at 0). This upper limit 
increases up to $\sim 50$ eV for $\sigma$ = 500 eV. These results are  
similar to the findings of Sakano et al. (1999) who
recently reanalyzed all the ASCA data on 1E~1740.7--2942.

However, a single power law cannot provide an adequate
fit over the broad energy range covered by BeppoSAX. In fact the 
inclusion of the PDS data clearly shows the presence of a spectral 
turn over at higher energies (see Fig. 6).
Good fits could be obtained with a power law with an exponential cut-off
or with Comptonization models (see Table~3).

1E 1740.7-2942 is of particular interest since, 
beside being one of the few galactic objects with persistent hard 
X-ray emission extending above $\sim$100 keV,  
it has a  peculiar radio counterpart with double sided relativistic jets 
(Mirabel et al. 1992).
The observation of a transient broad spectral feature 
at  $\sim$480 keV (Bouchet et al. 1991),
suggested that 1E 1740.7-2942
might be related to the 511 keV line observed from the GC region. On the 
basis of these properties and of its hard X-ray spectrum   similar to that of Cyg ~X-1,  
1E 1740.7-2942 is generally considered a black hole candidate. 
The broad band spectrum measured with BeppoSAX is consistent with 
this interpretation.

\placefigure{fig6}

\subsection{A~1742--294}

A~1742--294, although sometimes referred to as a transient source,
has  always been detected in all the GC observations
since 1975 (see references in van Paradijs 1995).

It was observed twice in the 
outer part of the MECS field of view (Obs. n. 8 and 9). 
In the 1997 observation  A~1742--294 was very close to the
region of the field of view used for the on board calibration
sources. So we extracted the counts for the spectral
analysis from a small circular region with $2'$ radius.
In both   observations acceptable fits were 
obtained both with a power law and with 
a thermal bremsstrahlung (see Table~3). In  March 1998  
the source  was softer and  brighter (of 
roughly a factor of two). Moreover, some variability
on a time scale of   hours was observed during 
both observations (see the light curves in Fig. 7). 
  
We detected three type I X--ray bursts from  A~1742--294:
one on September 16 (20:04:04 UT) and two on March 31, 1998 
(9:44:58  UT and 15:01:20 UT).
The poor statistics hampers a detailed
study of the neutron star luminosity, radius and temperature 
variations during the bursts. 

\placefigure{fig7}

\subsection{SLX~1744--299 and SLX~1744--300}

SLX~1744--299, first discovered with the Spartan 1 experiment 
(Kawai et al. 1988), was subsequently resolved into two distinct
objects separated by $\sim2.5'$ 
with the SL2-XRT coded mask telescope 
(Skinner et al. 1987, 1990).  

We obtained a single observation pointed on these two sources  
in September 1997 (Obs. n.7). For the spectral analysis we used extraction 
radii of only $2'$, and the same background estimated from a circular
corona surrounding both sources. Though the results have to be taken
with some caution due to the unavoidable cross-contamination of the
two spectra, we can  safely conclude that SLX~1744--300 is fainter
and has a  slightly softer spectrum than SLX~1744--299.
However, contrary to Predehl et al. (1995), 
we cannot claim a significant difference in their absorption column 
densities (see results in Table~3).

Though both SLX~1744--299 and SLX~1744--300 have been reported to
emit  X--ray bursts (Skinner et al. 1987, 1990; Sunyaev et al. 1991b), none 
was detected during our observation.

\subsection{XTE~J1748--288}

This transient source was discovered with the RossiXTE satellite in June
1998 (Smith, Levine \& Wood 1998). Its hard X--ray spectrum and its
association 
with a variable radio source with evidence for the presence of jets
establish XTE~J1748--288 as a possible black hole 
candidate 
(Strohmayer \& Marshall 1998, Hjellming et al. 1998, Fender \& Stappers
1998).

We detected XTE~J1748--288 in the August 1988 observation (Obs. n.11). 
Its spectrum, extracted from a $2'$ radius region to avoid the
problems due to the poorly known additional absorption   from
the strongback support of the MECS detector entrance window, 
could  be fit  equally well by a hard power law
(photon index $\sim$1.5) or by a thermal bremsstrahlung with kT$\sim$40
keV.
The flux  was $\sim$10$^{-10}$ erg~cm$^{-2}$ s$^{-1}$
(2--10 keV, corrected for the absorption).
The light curve of the XTE~J1748--288 outburst 
obtained with the RossiXTE ASM is shown in Fig.~8.
The date of our observation and the measured flux
are indicated by the dashed lines, showing that we obtained a
positive detection of XTE~J1748--288 when it was well below the ASM
sensitivity threshold.  
It is interesting to note that the flux measured with
$BeppoSAX$ is consistent with the extrapolation of
the   decaying light curve  observed with the  ASM
until about mid-July. 
This suggests that the nearly exponential decay, with e-folding
time $\sim$20 days, lasted until the date of our observation
(August 26th).

\placefigure{fig8}

\subsection{SAX~J1747.0--2853}

The source SAX~J1747.0--2853, 
positionally consistent with the transient GX~0.2--0.2
observed in  1976 (Proctor, Skinner \& Willmore   1978), was discovered 
with the  $BeppoSAX$ Wide Field Camera (WFC) instrument 
during an outburst in March 1998 (in't Zand et 
al. 1998b, Bazzano et al. 1998).

\placefigure{fig9}

During our observation  pointed on  1E~1743.1--2843 (Obs. n.10),   
SAX~J1747.0--2853  was detected at an off-axis angle of $\sim13'$.
During this observation, performed about 20 days
after the WFC ones,  
a Type-I X--ray burst from this source  was seen (Sidoli et al. 1998b).
The burst, starting at 1:40 UT of 1998 April 15, 
was also visible at energies above  10 keV with the PDS instrument. 
The burst light curve in different energy ranges is shown in Fig.~9,
where we also present the variations of the spectral fit parameters
during the burst (see  Sidoli et al. 1998b for details). 
The spectral softening  clearly indicates that the burst is of Type I,
confirming the LMXRB nature of this source.   
Assuming an Eddington luminosity at the burst
peak, we obtain a distance of $\sim$10 kpc,
and a blackbody radius consistent with the expectations for  a neutron star.

\subsection{KS~1741--293}
 
During our observation pointed on the molecular cloud SgrC performed 
in March 1998 (Obs. n.9), we detected a source positionally coincident with 
the transient burster KS1741-293 (in't Zand et al. 1998c).

Since the source was located close to the strongback support
of the MECS detector entrance window, we used a small
extraction radius ($2'$) for the spectral analysis.
The background was estimated locally from a source free region 
of the same observation. 
The spectrum is relatively soft and could be described equally well
by a power law and  a thermal bremsstrahlung  (see Table~3),
giving in both cases a high column density   
$N_{H}\sim$2$\times10^{23}$ cm$^{-2}$, and a 
luminosity   L$_{X}\sim10^{36}$ erg~s$^{-1}$
(2--10 keV, corrected for absorption).
Slightly lower column densities and luminosity were obtained with
a blackbody spectrum (see Table~3).

The same region of sky was observed by $BeppoSAX$ in September 1997 (Obs. n.8)
and  KS~1741--293  was not detected, with an upper limit of 
L$_{X}$ $<$ 10$^{35}$ erg~s$^{-1}$.

\subsection{1E~1743.1--2843}

This bright and highly absorbed source has been repeatedly observed 
in several of our pointings. The best fit results 
reported in Table~3 refer to the
observation performed in April 1998 (Obs. n.11), in which 1E~1743.1--2843
was on-axis.
This source is probably a persistent LMXRB, though no X--ray
bursts have been detected so far. 
For a complete discussion of all the $BeppoSAX$ and
ASCA observations of  1E~1743.1--2843 see Cremonesi et al. (1999).

\subsection{The supernova remnant G0.9+0.1}

The discovery of X--ray emission from the center of the radio
supernova remnant G0.9+0.1 has been   reported by Mereghetti, Sidoli
\& Israel (1998). 
The location of G0.9+0.1 was imaged in April and September 1997, during
observations pointed on the Sgr~B2 molecular cloud.  The angular
resolution of the MECS at the off--axis location of G0.9+0.1 ($\sim14'$) 
only allowed to establish that the X--ray emission is associated with the
plerionic radio core of the supernova remnant and not with the surrounding
shell ($\sim8'$ diameter, Helfand \& Becker 1987).  However, it was not
possible to discriminate between a point source (i.e. a neutron star) and
a synchrotron nebula of a few arcmin extent.

We have performed a new   spectral analysis based on the improved 
calibration results that are now available. 
The derived spectral parameters  are within the  
uncertainties of those previously reported
(Mereghetti, Sidoli \& Israel  1998), but they 
are more precise.

The best fit power--law spectrum has a photon index
$\Gamma=2.5$, $N_H=2.5\times 10^{23}$\cmdue, and flux $F = 2\times
10^{-11}$ erg cm$^{-2}$ s$^{-1}$ (2--10 keV, corrected for the
absorption). Equally good fits were also obtained with blackbody
and thermal bremsstrahlung spectra (see Table~3), while a
Raymond-Smith thermal plasma model with abundances fixed at the solar
values gave a worse result.

The $BeppoSAX$ discovery of X--ray emission from the central region of
G0.9+0.1 confirms its plerionic morphology derived from the radio
observations.  The presence of a young, energetic pulsar in G0.9+0.1 could
also explain part of the high energy gamma-ray emission observed with
EGRET from the region of the Galactic Canter (Mayer--Hasselwander et al.
1998).

\subsection{G359.23--0.92 (The Mouse)}

The radio source  G359.23--0.92, also known as the Mouse,
belongs to a small class of
radio nebulae characterized by an axially symmetric morphology
(Yusef-Zadeh \& Bally 1987). It is believed that these sources are produced by
relativistic particles ejected by a compact object (either a 
pulsar or a binary system) moving with high speed along the axis
of symmetry (Shaver et al. 1985, Helfand \& Becker 1985).
Predehl \& Kulkarni (1995) using the ROSAT PSPC instrument
discovered  X--ray  emission associated with the ``head'' of the Mouse,
located only $2.3'$ northeast of SLX~1744--299.
Since the MECS data of  SLX~1744--299/300   showed some evidence
for an elongation in the direction of G359.23--0.92,
we performed the same procedure described
in section 4.2 for AX~J1745.6--2901.
The resulting 6--10 keV image, shown in Fig.~10, confirms the
MECS detection of  G359.23--0.92.
 
\placefigure{fig10}

To estimate the spectral parameters of  G359.23--0.92 we used a small extraction radius ($2'$) and  
measured  the background (which is dominated by the counts of the
nearby sources)   from a  similar region at the same distance from SLX~1744--299.
A power law fit   gave $\Gamma\sim$ 1.9--2.3, a 
column density of $4-6\times$10$^{22}$~cm$^{-2}$ and an unabsorbed flux in the 2--10 keV
band of the order of $3\times$10$^{-11}$ erg~cm$^{-2}$ s$^{-1}$.

This flux is higher than the value expected based on the ROSAT results
at energies below 2.4 keV (Predehl \& Kulkarni 1995), however  this discrepancy
might be due to different reasons other than source variability.
First,   the flux
reported by these authors was based on the assumption of a value of only
$2\times10^{22}$~cm$^{-2}$ for the column density: we found that for our N$_{H}$ value
the MECS and ROSAT PSPC count rate are consistent with the same flux level.
Another possibility is that we underestimated the contamination from SLX~1744--299/300.

In any case, the detection at energies greater than 6 keV is statistically significant and supports
a non-thermal origin for the X--ray emission from the head of the ``Mouse''.

\section{Discussion}

Our $BeppoSAX$ survey of the inner  4 degrees of the galactic
plane shows that a few   persistently bright sources characterize 
the constellation of
X--ray sources near the GC.  Their properties indicate
that they consist of compact accreting objects similar to the classical
accretion powered sources found elsewhere in the Galaxy.
Both systems containing neutron stars (e.g. SLX~1744--299/300, A~1742--294)
and likely black holes (1E~1740.7--2942 ) are present.

However, these persistently bright objects represent only a small fraction of
the X--ray sources in this region. Three transient
sources (SAX~J1747.0--2853, KS~1741--293 and XTE~J1748-288) were
detected in our survey, but it is known from previous observations
that many more are present. 
We therefore confirm   
the clustering of X--ray sources towards the GC 
already noted by several authors.

It is likely that on the large scale ($\sim$ degrees) this simply
reflect the general mass distribution enhancement, as found 
by Grebenev,  Pavlinsky \& Sunyaev (1996) who considered only the  
hard X--ray sources detected with the  ART-P coded mask telescope. 
On the other hand, considering the distribution of all the sources
reported within several arcminutes from SgrA* there is
marginal evidence of an additional source population 
(Skinner 1993).  However, this conclusion might be biased by the
different sensitivity and angular resolution of the many observations
of this field. Though it is obvious that the regions closer to the 
center have been observed more deeply, the effect of this bias is 
difficult to quantify.
We finally note that the case of the three sources in the direction of
SLX J1744--299/300 (at b = --1.5$\deg$ ) is maybe more surprising that the presence of 4-5
sources close to SgrA*. While G359.23--0.92 is very likely a foreground
object, the two X--ray bursters SLX J1744--299 and SLX J1744--300
could be at the same distance and therefore somehow related 
(e.g. members of a star cluster). Observations in X-rays with better 
angular resolution and deep searches in the infrared are needed
to study these objects.

\placefigure{fig11}

In Fig.~11 we have plotted the column density and power law photon index 
(in the 2--10 keV range) for all the sources that we have observed 
(to compare the spectra of the different sources,
we have used a power law also in a few cases in which other models
provided better fits). 
It is clear that sources containing neutron stars, as indicated
by the presence of Type I bursts, have softer spectra than the 
black hole candidates. Based on this spectral dichotomy, 
it is likely that also 1E~1743.1-2843
contain a neutron star.
Fig.~11 also shows that the  column densities of these sources 
vary over a    factor $\sim$4 range. 
This  is probably   related to
the clumpiness of the molecular clouds in this region rather than to
a corresponding spread in the source distances.

\subsection{Sgr A* and the galactic center X--ray source(s)}

At least four point sources (Predehl \& Trumper 1994, Maeda et al. 1996),
maybe five (Kennea \&  Skinner 1996) and possibly more could
contribute to the X--ray flux detected with $BeppoSAX$
when pointing at the GC, 
in addition to the diffuse emission
that permeates the region (Maeda \& Koyama 1996).
However, as shown in Fig~2, the MECS source   position is closer to 
the ROSAT counterpart
of SgrA*  than to the other two ROSAT sources (PT6 and PT8).  

The emission lines detected in the MECS spectrum within $2'$ from the GC 
clearly indicate a substantial 
contribution from the thermal diffuse emission associated to the 
SgrA West and SgrA East regions (Yusef-Zadeh et al. 1997). 
Unfortunately, this contribution is difficult to quantify, and the limited angular
resolution does not allow to measure the flux to be ascribed to SgrA* itself.

We can place a very conservative upper limit to the luminosity of SgrA* by assuming
that it is the only point source and that the diffuse emission at its position
has the same surface brightness of the  annular region ($6'-8'$)
described above. In this case, using the flux reported in Table~3,
we obtain a luminosity of $\sim3\times10^{35}$~ergs~s$^{-1}$ 
(2--10 keV, corrected for the absorption).
However, this is probably an overestimate of the true luminosity of SgrA* for the fact
that the diffuse emission has a surface brightness distribution that increases
toward the GC and because 
there might be a contribution from the other point sources.
 
A more realistic upper limit to the X--ray emission  from   SgrA*
can be placed if we consider an  X--ray background rising towards the GC: 
we measured the surface brightness  from three concentric 
annular regions ($2'-4'$, $4'-6'$ and  $6'-8'$)
and extrapolated it at our source position. 
In this case we find a new background value, four times higher than the previous one
and the 
new   flux for the point source  is   
$\sim1.5\times 10^{-11}$ ergs \cmdue s$^{-1}$, that translates into   
a luminosity  at 8.5 kpc of $\sim10^{35}$~ergs~s$^{-1}$. 
This upper limit is similar to   the value reported 
by the ASCA team (Maeda et al., 1998) and is
 well below the Eddington luminosity
for a super--massive black hole. 

Several authors tried to explain this high energy 
underluminosity of SgrA* with different models for the accretion, 
from a spherical accretion
 (Melia 1992)
to the advection dominated accretion flow 
(Narayan, Yi \& Mahadevan 1995; Narayan et al. 1998). These models 
have still some problems in the radio and in the $\gamma$--ray bands  (EGRET source
2EG~1746--2852), where   substantial
excesses  are present. Several solution has been 
proposed: the inclusion of an emission process 
associated with protons (Mahadevan et al. 1998) is able to solve the problem
of the radio excess, while the EGRET source is considered as a gamma--ray 
 upper limit due to the poor spatial resolution. Another possible explanation is
 that the EGRET source is not related with SgrA*, but with the 
 SgrA East supernova remnant shell (Melia et al. 1998). 
 
 The XMM and AXAF observatories with    their superior 
 imaging capabilities will substantially contribute 
 to solve the problem of the high energy emission from SgrA* and
 put more stringent contraints to the accretion models.

\acknowledgments
We thank  Silvano Molendi for useful discussions and help with the data analysis.

\clearpage
 
\begin{deluxetable}{cclcl}
\footnotesize
\tablecaption{BeppoSAX observations summary. \label{tbl-1}}
\tablewidth{0pc}
\tablehead{
\colhead{Obs.} & \colhead{Pointing Direction}   & \colhead{Observation Date}   & \colhead{MECS Exposure} & 
\colhead{Main}    \nl 
\colhead{n.}     & \colhead{R.A. and Dec. (J2000)}  & 
\colhead{Start \& Stop time (UT)}  & \colhead{Time (ks)} & \colhead{target}}

\startdata
  1 &  17 43 54.8  --29 44 42	& 1996 Sep   03 10:56 - 03 16:30 & 7   & 1E 1740.7--2942 \nl
  2 &  17 47 20.0  --28 24 02	& 1997 Apr   05 08:57 - 06 09:33 & 49  & Sgr B2            \nl
  3 &  17 45 40.3  --29 00 23	& 1997 Aug   24 06:01 - 26 08:41 & 100 & Sgr A*            \nl
  4 &  17 47 20.0  --28 24 02	& 1997 Sep   03 05:18 - 04 09:36 & 51  & Sgr B2           \nl
  5 &  17 48 39.2  --28 05 56	& 1997 Sep   04 22:45 - 06 03:22 & 49  & Sgr D            \nl
  6 &  17 43 54.8  --29 44 42	& 1997 Sep   07 05:45 - 08 06:57 & 43  & 1E 1740.7--2942  \nl
  7 &  17 47 25.8  --30 01 01	& 1997 Sep   10 14:07 - 11 18:36 & 21  & SLX 1744--299    \nl
  8 &  17 44 41.3  --29 28 14	& 1997 Sep   16 11:30 - 16 20:47 & 15  & Sgr C            \nl
  9 &  17 44 41.3  --29 28 14	& 1998 Mar   31 06:47 - 31 20:54 & 25  & Sgr C           \nl
 10 &  17 46 19.9  --28 44 00 	& 1998 Apr   13 13:36 - 15 05:51 & 71  & 1E 1743.1--2843  \nl
 11 &  17 46 50.2  --28 33 04	& 1998 Aug   26 16:34 - 28 16:30 & 79  & Sgr B1            \nl
\enddata
\end{deluxetable}

\clearpage

\begin{deluxetable}{llllll}
\footnotesize
\tablecaption{X--ray sources in the observed region. \label{tbl-2}}
\tablewidth{0pc}
\tablehead{
\colhead{Source} & \colhead{RA (J2000)} & \colhead{Dec (J2000)} & \colhead{Error} & \colhead{Notes} & \colhead{Ref.\tablenotemark{a} } }

\startdata
XTE~J1748--288		&17 48 08	&-28 28 48	& $1'$	& Transient	&1	   	\nl
SLX~1744--299		&17 47 27	&-29 59 55 	& $1'$	& Burster 	&2	     	\nl
SLX~1744--300		&17 47 27	&-30 02 15 	& $1'$	& Burster	&2 	   	\nl
G0.9+0.1 		&17 47 21 	&-28 09 22  	& 	& SNR	&3    \nl
G359.2--0.8 (The Mouse) &17 47 15.2	&-29 58 00  	& $8''$	&  		&4	   	\nl
SAX~J1747.0--2853	&17 47 0	&-28 52 12	& $1'$	& Transient/Burster	&5 \nl
1E~1743.1--2843		&17 46 19.5	&-28 53 43	& $1'$	&  		&6	 \nl
A~1742--294		&17 46 05	&-29 30 54 	& $1'$	& Burster	&7	   \nl
RXJ~1745.7--2858	&17 45 45.5 	&-28 58 17 	& $20''$& Star? (PT6)   & 8 \nl
1E~1742.5--2859		&17 45 40.7 	&-29 00 10 	& $1'$	& SgrA*?   & 6 \nl
RXJ~1745.6--2900 	&17 45 40.4 	&-29 00 23 	& $20''$& SgrA*? (PT7) 	&8	 	\nl
AX~J1745.6--2901	&17 45 36 	&-29 01 34	& $25''$	& Transient/Burster?	&  9  \nl
RXJ~1745.5--2859	&17 45 32 	&-28 59 29 	& $20''$&  PT8	&  8  \nl
A~1742--289		&17 45 37.3	&-29 01 04.8    & $2.8''$& Transient &  10 \nl
KS~1741--293		&17 44 49.2	&-29 21 06.3	& $1'$	& Transient/Burster	&11	 \nl
1E~1740.7-2942  	&17 43 54.8	&-29 44 42.8 	& $1''$	& Black hole candidate	&12	\nl
 \tablenotetext{a}{References are: 1=Strohmayer \& Marshall 1998, 2=Skinner et al. 1990, 
3=Helfand \& Becker 1987,  4=Predehl \& Kulkarni 1995, 5=Bazzano et al. 1998, 6=Watson et al. 1981, 
7=Pavlinsky, Grebenev \& Sunyaev 1994, 8=Predehl \& Trumper 1994, 9=Maeda et al. 1996, 10=Davies et al. 1976, 
11=In't Zand et al. 1990, 12=Mirabel et al. 1992}
\enddata
\end{deluxetable}

\clearpage

\begin{deluxetable}{llclcl}
\footnotesize
\tablewidth{0pc}
\tablecaption{Results of the Spectral Fits  (errors are 90\% confidence level). \label{tbl-3}}
\tablehead{
\colhead{Source}         & \colhead{Model\tablenotemark{a}}      &
\colhead{Column density}          & \colhead{Parameter\tablenotemark{b}}  &
\colhead{Red.$\chi^2$ (d.o.f.)}          & \colhead{Flux\tablenotemark{c}}   \nl
\colhead{(data)  } & \colhead{ } & \colhead{($10^{22}$ cm$^{-2}$)}  &
\colhead{ } & \colhead{ } &
\colhead{ }}
  
\startdata
Galactic center	&PL+line(6.7)		& $8.3\pm{0.5}$		& $\Gamma=2.6\pm{0.1}$  & $1.12$ (284)	&  $4.0\pm{0.2}$	\nl
(MECS)		&			&			& $EW(6.7)=1.2  $	&		&   			\nl
		&Brems+line(6.7)		& $6.6\pm{0.3}$		& $T = 5.1\pm{0.5}$	& $1.20$ (284)	&  $3.3\pm{0.2}$	\nl
		&			&			& $EW(6.7)=1.2  $	&		&   			\nl
 		&M 			& $7.0\pm{0.3}$		& $T = 4.1\pm{0.3}$	& $1.16$ (288)	&  $3.5\pm{0.2}$	\nl
		&M+M 		& $11.3\pm{1.2}$	& $T_{M1} = 0.80^{+0.18}_{-0.08}  $& $1.00$ (285)&   $6.0^{+0.6}_{-1.0} $\nl
		&			&			& $T_{M2} =4.9\pm{0.4}$&		&   			\nl
		&M+PL+line(6.7)	& $10\pm{1}$			& $T_{M}=1.3^{+0.2}_{-0.6}$ &$0.97$ (281)&   $5.5\pm{1}$	\nl
		&			&			& $\Gamma=1.7\pm{1}$&		&   			\nl
		&			&			& $EW(6.7)=0.7-1  $	&	&   \nl		
\hline
1E~1740.7--2942	&PL			& $14.7\pm{0.4}$	& $\Gamma=1.52\pm{0.04}$& $1.10$ (180)	&  $49.5\pm{1.0}$ \nl
(MECS)		&Brems			& $14.1\pm{0.3}$	& $T=34.5\pm{6.0}$	& $1.12$ (180)	&  $47.4\pm{0.4}$ \nl
\cline{2-6}
(LECS+MECS 	&PL			& $14.5\pm{0.3}	$	& $\Gamma=1.45\pm{0.3}$	& $2.84$ (283)	&  $49\pm{1}$ \nl
+PDS)		&Brems 			& $12.9\pm{0.2}$	& $T=174\pm{6}$		& $1.80$ (283)	&  $45\pm{0.5}$  \nl
		&CompST			& $14.6\pm{0.2}$	& $T=24\pm{1}$		& $1.30$ (282)	&  $49\pm{0.5}$ \nl
		& 			&			& $\tau=5.5\pm{0.1}$	&	&   \nl
		&PL+E$_{c}$ 		& $13.6\pm{0.2}$	& $\Gamma=1.36\pm{0.02}$	& $1.24$ (281)	&  $46.7\pm{0.5}$ \nl
		&			&			& E$_{c}=52\pm{4}$  &  &   \nl
		&			&			& E$_{fold}=105\pm{10}$  &	  & \nl		
\hline 
A~1742--294	&PL			& $6.5\pm{0.5}$		& $\Gamma=1.72\pm{0.11}$& $0.995$ (238)	&  $30\pm{1}$ \nl
(MECS-Obs.9)	&Brems			& $5.9\pm{0.4}$		& $T=16.3^{+5.1}_{-3.6}$& $0.986$ (238)	&  $29\pm{1}$ \nl
\cline{2-6}
     		&PL			& $6.9\pm{0.3}$		& $\Gamma=1.97\pm{0.05}$& $0.99$ (323)	&  $63\pm{1}$ \nl
(MECS-Obs.10)	&Brems			& $6.0\pm{0.2}$		& $T=10.3\pm{1.0}$	& $1.01$ (323)	&  $58\pm{1}$ \nl
\hline
SLX~1744--299	&PL			& $5.1\pm{0.2}$		& $\Gamma=2.1\pm{0.1}$	& $0.92$ (364)	&  $20\pm{0.4}$ \nl
(MECS)		&Brems			& $4.2\pm{0.2}$	  	& $T=8.9\pm{0.6}$	& $0.99$ (364)	&  $18\pm{0.4}$ \nl
\hline
SLX~1744--300	&PL			& $5.3\pm{0.2}$		& $\Gamma=2.2\pm{0.1}$	& $1.06$ (342)	&  $12\pm{0.3}$ \nl
(MECS)		&Brems			& $4.2\pm{0.2}$	  	& $T=7.7\pm{0.5}$	& $1.09$ (342)	&  $11\pm{0.3}$ \nl
\hline 
XTE~J1748--288	&PL			& $7.3\pm{0.5}$		& $\Gamma=1.46\pm{0.1}$	& $0.922$ (163)	&  $11.2\pm{0.3}$ \nl
(MECS)		&Brems			& $6.9\pm{0.4}$	  	& $T=40^{+24}_{-11}   $	& $0.93$ (163)	&  $10.9\pm{0.3}$ \nl
\hline 
SAX~J1747.0--2853&Brems			& $8.3^{+0.6}_{-0.3}$	& $T=6.1^{+0.9}_{-0.7}$	& $1.01$ (138)	&  $4.0^{+0.2}_{-0.3}$ \nl
(MECS)		&			&			&			&		&   	\nl
\hline 
KS~1741--293	&PL			& $20\pm{2}$		& $\Gamma=2\pm{0.2}$	& $0.992$ (213)	&  $14.5^{+2.0}_{-0.6}$  \nl
(MECS)		&Brems			& $18\pm{1}$		& $T = 11\pm{3}$	& $0.917$ (213)	&  $12.5^{+1.2}_{-0.5}$ \nl
		&BB			& $12\pm{2}$		& $T = 1.8\pm{3}$	& $0.966$ (213)	&  $8.5\pm{0.4}$ \nl	 
\hline
1E~1743.1--2843	&BB			& $13\pm{1}$	& $T = 1.8\pm{1}$& $1.01$ (178)&   $16.5\pm{0.5}$	\nl
(MECS)		&			&			&			&		&   	\nl
\hline
G0.9+0.1	&PL			& $25^{+17}_{-10}$	& $\Gamma=2.5^{+1.5}_{-0.8}$& $0.72$ (28)       &   $2.0^{+4.0}_{-1.0}$	\nl
(MECS)		&Brems			& $23^{+12}_{-8}$	& $T=6^{+13}_{-3}   $	   &	$0.72$ (28)	&   $1.5^{+0.2}_{-0.5}$	\nl
	 	&BB			& $16^{+10}_{-10}$	& $T=1.6^{+0.6}_{-0.5}   $ &	$0.72$ (28)	&   $0.9^{+0.2}_{-0.4}$	\nl
\hline
G359.23-0.92	&PL			& $4-6$	& $\Gamma=1.9-2.3$&  &   $\sim3$	\nl
(MECS)		&			&			&			&		&   	\nl
\tablenotetext{a}{PL=Power Law,PL+E$_{c}$= Power law plus high energy cutoff (E$_{c}$=cutoff energy in keV;
E$_{fold}$=e-folding energy in keV), BB=Blackbody, 
Brems=Thermal Bremsstrahlung, CompST=Comptonization model
  (Sunyaev \& Titarchuk 1980), line=gaussian line, M=MEKAL model in XSPEC.} \nl
\tablenotetext{b}{All the temperatures and energies are in keV, $\Gamma$=photon index.} \nl
\tablenotetext{c}{Unabsorbed fluxes (2--10 keV) are in units of $10^{-11}$ ergs cm$^{-2}$ s$^{-1}$}
\enddata
\end{deluxetable}

\clearpage

\figcaption[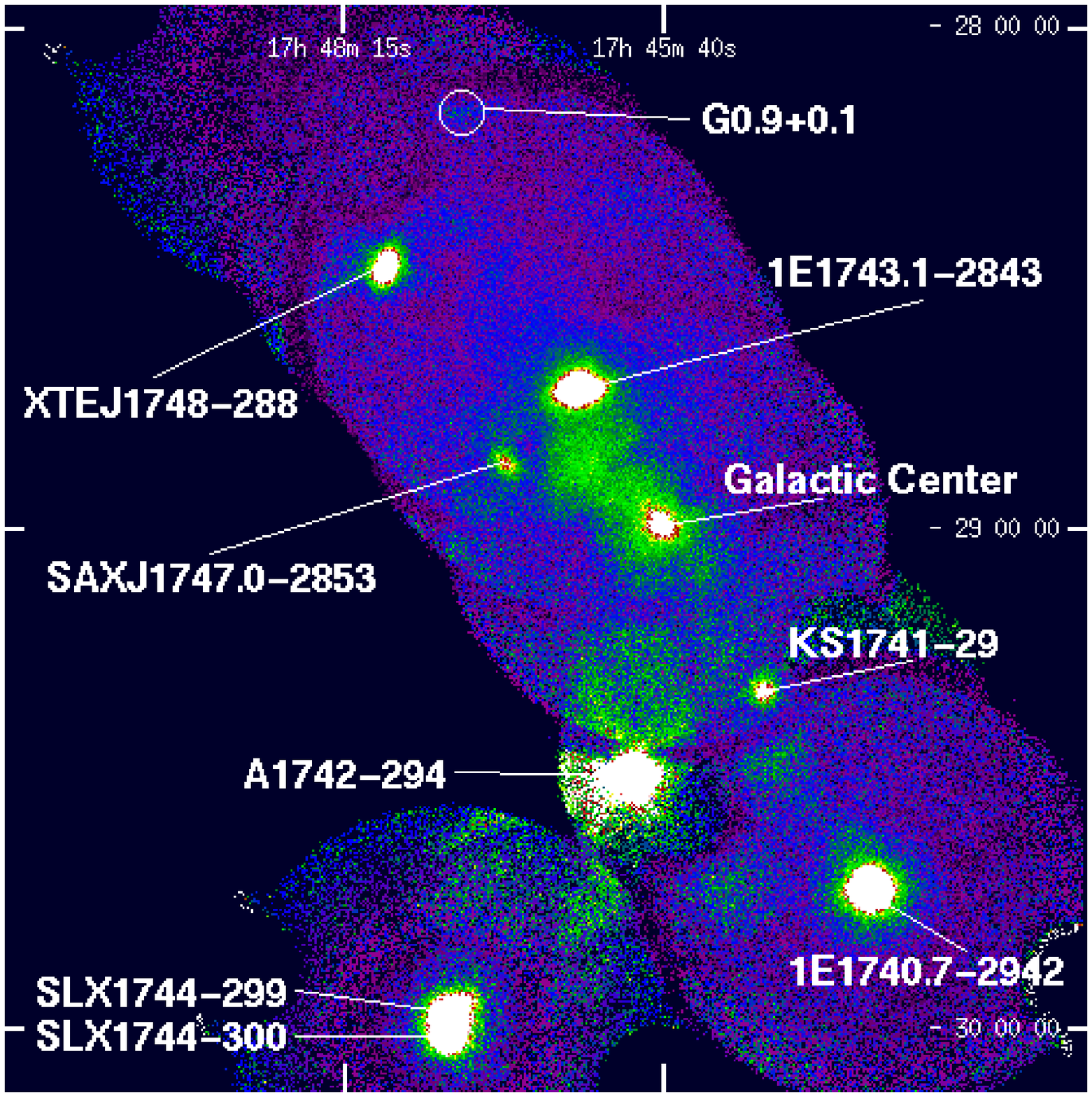]{Mosaic of the Galactic Center 
region images obtained with the MECS3 
instrument in the 2--10 keV energy range. North is to the top, East to
the left. Coordinates are for the J2000 equinox. 
The image has been corrected for the vignetting and 
for the differences in the
relative exposure lengths of the various observations. \label{fig1}}

\figcaption[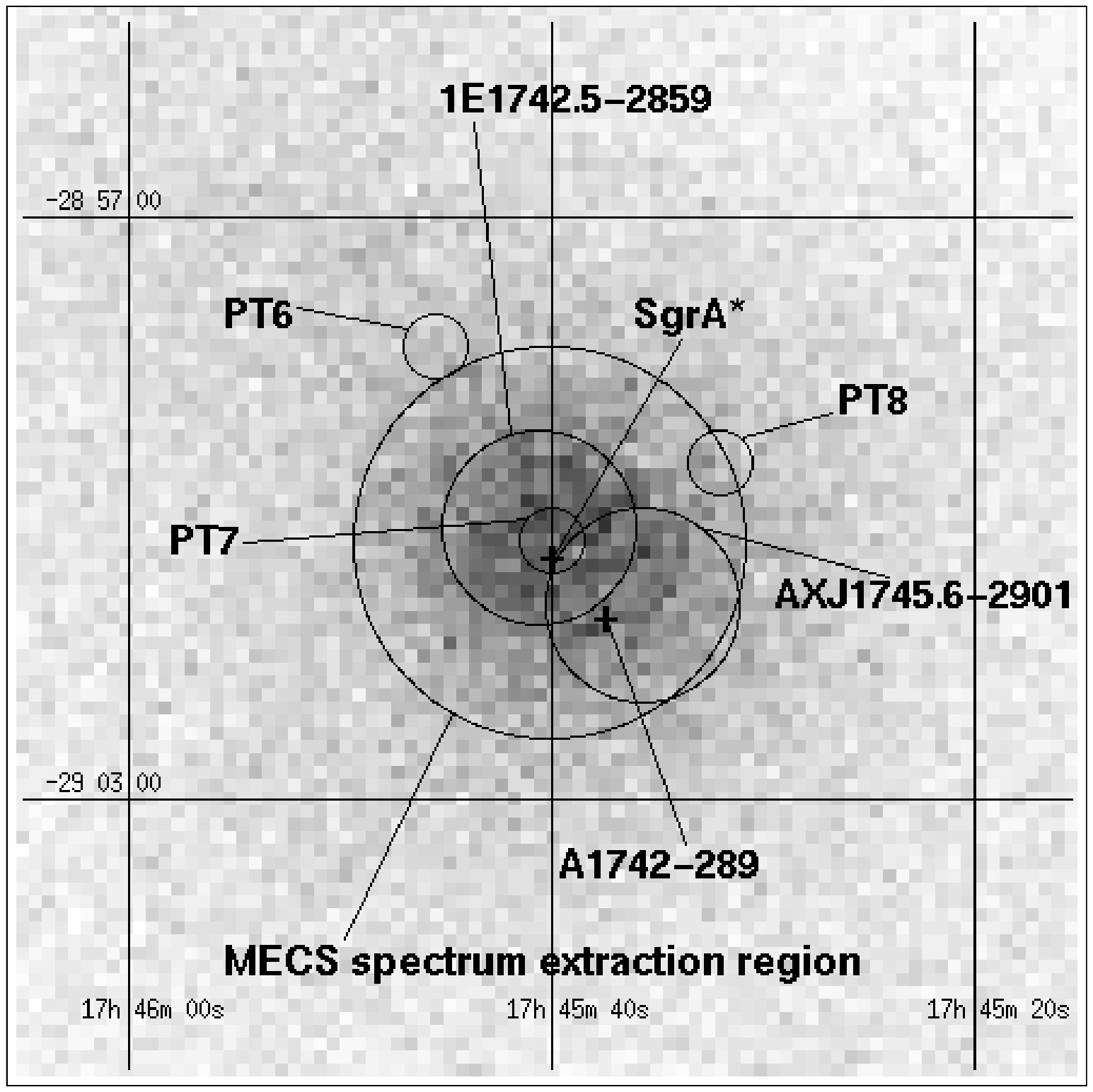]{Positions of the X--ray sources in the vicinity of SgrA* superimposed
on the MECS 2--10 keV image. The error circles have radii of $20''$  
for the ROSAT sources (PT, from Predehl \& Trumper 1994), 
$1'$ for AX~J1745.6--2901 (Maeda et al. 1996)   and $1'$ 
for 1E~1742.5--2859 (Watson et al. 1981). The two crosses mark 
the accurate positions, obtained with   radio observations,
of SgrA* ($R.A.=17h~45m~40.131s, 
Dec.=-29\deg~00'~27.5''$, Menten et al. (1997) ) 
and A~1742--289 (Davies et al. 1976). 
The large circle
($2'$ radius) corresponds to the extraction region 
of the counts used in the spectral analysis. 
All the coordinates are for the J2000 equinox. \label{fig2}}

\figcaption[fig3.ps]{Radial profile of the 
X-ray emission from the GC region  in the 2--5 keV  range (upper panel) 
and  in the 5--10 keV range (lower panel).  
The data points represent the surface brightness measured in concentric
rings centered at the position of SgrA*, while the solid
lines show for comparison
the profiles expected in the case of a single unresolved point source.
Both curves are background subtracted. The expected profiles have
been normalized to yield the same number of 
counts within 10$'$ as the measured data.
Independently 
of the  relative normalization, 
the different slope of the two profiles   demonstrates that a single
point   source at the GC cannot account
for all the observed emission.
   \label{fig3}}

\figcaption[fig4.ps]{ 
Spectrum of the X-ray emission within 2$'$ of SgrA* fitted with
a thermal plasma model with kT=4 keV (both MECS). The residuals at energies
around $\sim$2.4 keV suggest the presence of a multitemperature
plasma.   \label{fig4}}

\figcaption[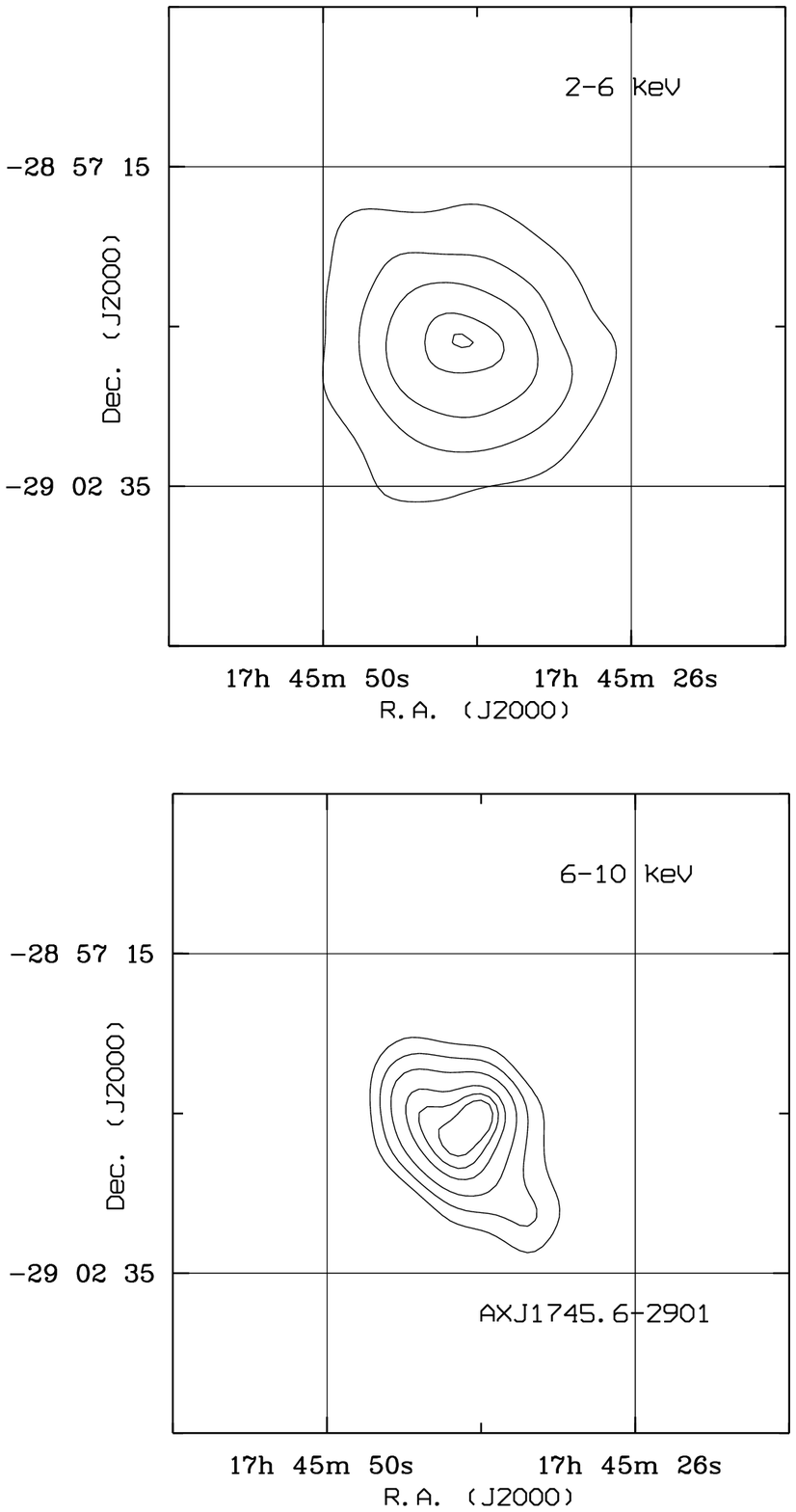]{ 
Images of the SgrA* region obtained with a single    MECS  unit (M3).
The top and bottom panels correspond respectively to the soft (2--6 keV) 
and hard (6--10 keV) energy range.  \label{fig5}}

\figcaption[fig6.ps]{Spectrum of the black hole 
candidate 1E 1740.7-2942 fitted with a cut-off power law
(see Table~3 for the parameters). \label{fig6}}

\figcaption[fig7.ps]{ Background subtracted light curves of 
A 1742-294, binned in 5000 s long time intervals. \label{fig7}}
 
\figcaption[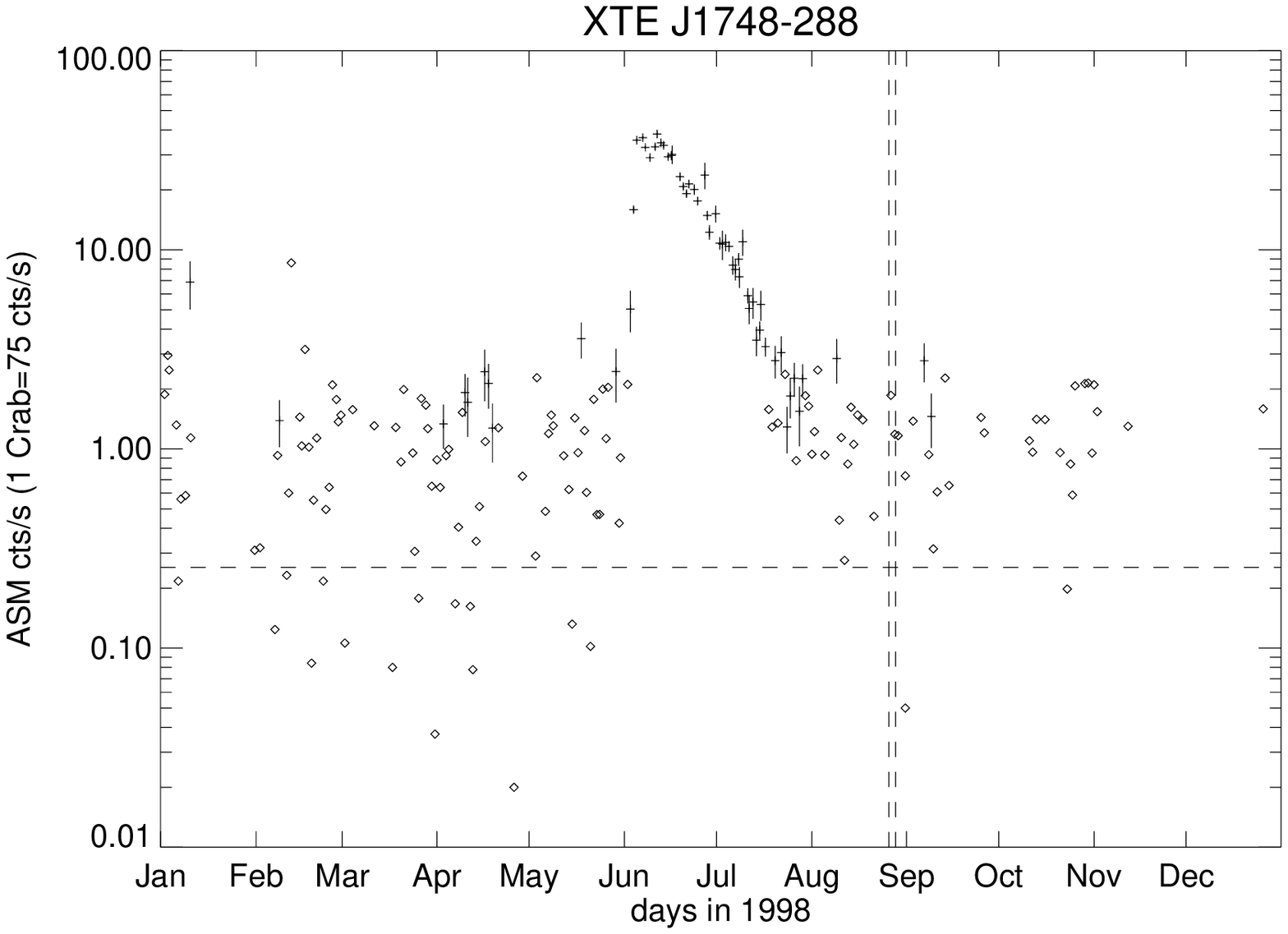]{
Detail of the 1998 outburst of XTE J1748-288 as observed by the XTE ASM. The
ASM detections at more than 3 sigma are plotted as error bars, while the
remaining ASM points are shown as diamonds. The vertical dashed lines indicate
the epochs of start and end of our observation (n. 11), while the horizontal
dashed line indicates the flux level measured by BeppoSAX. 
Note that the  BeppoSAX flux measurement represents a significant
detection of XTE J1748-288 at a level well below the ASM sensitivity 
threshold and consistent with an extrapolation of the ASM light curve.
 \label{fig8}}

\figcaption[fig9.eps]{
Left panel: light curve of the burst from SAX J1747.0--2853 in different 
energy ranges. The dotted and dashed lines indicate the level of
the persistent emission in the contiguous time interval respectively
before and after the burst. The burst is well detected also in the 
PDS instrument above $\sim$12 keV.
Right panel: Results obtained by fitting the burst emission in different
time intervals with a blackbody model.   \label{fig9}}

\figcaption[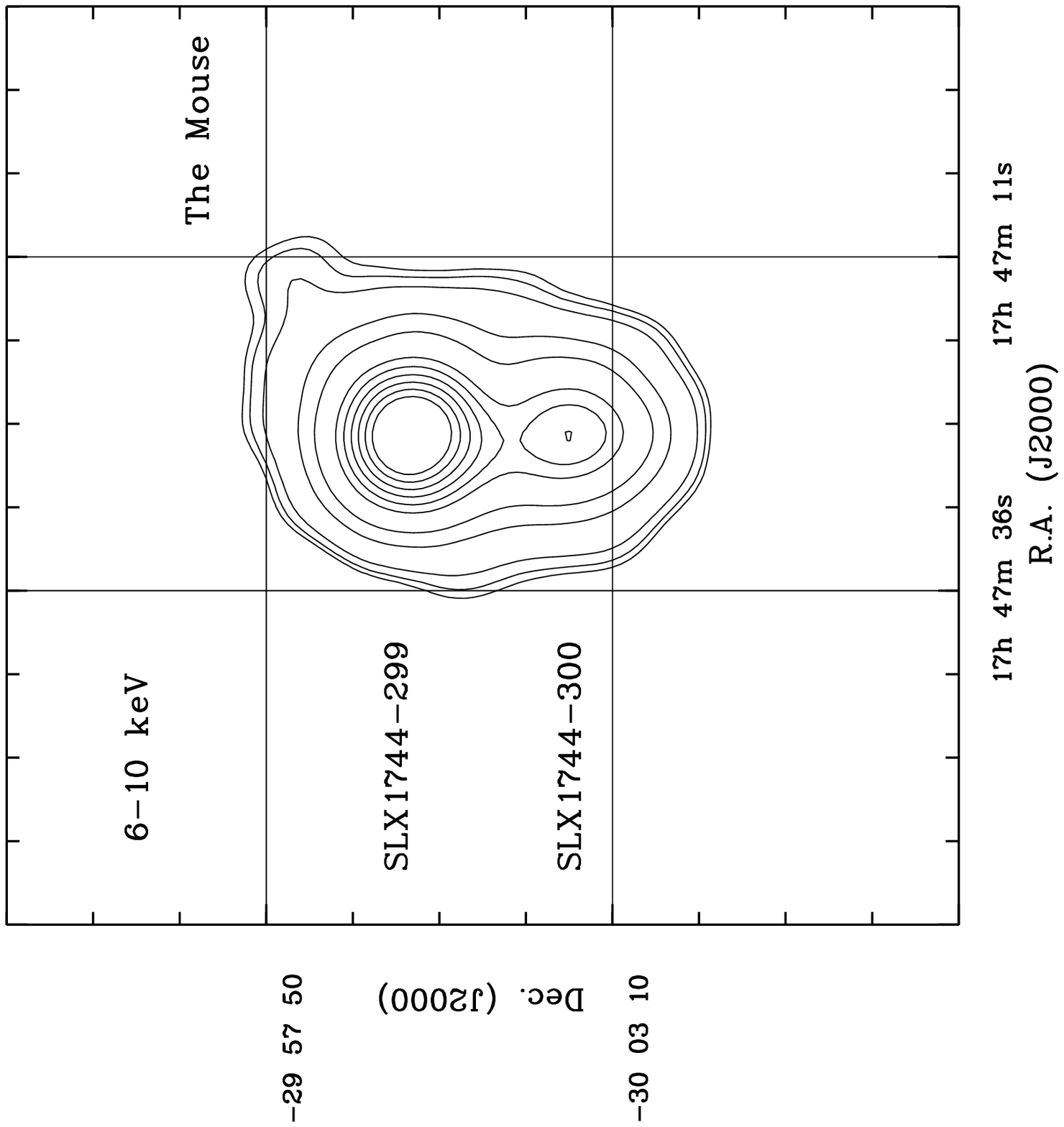]{MECS Image (M3 unit) of the 
region containing SLX~1744--299, SLX~1744--300 and
the ``Mouse'' in the 6-10 keV energy range.  \label{fig10}}

\figcaption[fig11.ps]{Absorbing column density 
and 2--10 keV spectral slope (power law photon index) 
for the sources in the GC region. The triangles mark 
the black hole candidates, 
while sources that showed Type I X-ray bursts are 
indicated with squares. \label{fig11}}

\end{document}